\documentclass[prl,twocolumn,showpacs,amsmath,amssymb,superscriptaddress,floatfix]{revtex4-1}
\usepackage[pdftex]{graphicx,hyperref}
\hypersetup{colorlinks = true, urlcolor = blue, linkcolor = blue, citecolor = blue}

\newcommand{\nc}{\newcommand}

\nc{\be}{\begin{equation}} \nc{\ee}{\end{equation}}
\nc{\bea}{\begin{eqnarray}} \nc{\eea}{\end{eqnarray}}
\nc{\bean}{\begin{eqnarray*}} \nc{\eean}{\end{eqnarray*}}
\nc{\dg}{\dagger}
\nc{\ua}{\uparrow} \nc{\da}{\downarrow}
\nc{\lag}{\langle} \nc{\rag}{\rangle}

\begin{document}

\title{Helical Hinge Majoranas in Iron-Based Superconductors}
\author{Rui-Xing Zhang}
\email{ruixing@umd.edu}
\author{William S. Cole}
\email{wcole1@umd.edu}
\author{S. Das Sarma}
\affiliation{Condensed Matter Theory Center and Joint Quantum Institute, Department of Physics, University of Maryland, College Park, Maryland 20742-4111, USA}

\begin{abstract}
Motivated by recent experiments on FeTe$_{1-x}$Se$_{x}$, we construct an explicit minimal model of an iron-based superconductor with band inversion at the $Z$ point and non-topological bulk $s_{\pm}$ pairing. While there has been considerable interest in Majorana zero modes localized at vortices in such systems, we find that our model - {\it without} any vortices - intrinsically supports 1D helical Majorana modes localized at the hinges between (001) and (100) or (010) surfaces, suggesting that this is a viable platform for observing ``higher-order'' topological superconductivity. We provide a general theory for these hinge modes and discuss their stability and experimental
manifestation. Our work indicates the possible experimental observability of hinge Majoranas in iron-based topological superconductors.
\end{abstract}
\date{\today}
\maketitle

{\it Introduction} - In the decade since their initial discovery, iron-based superconductors (FeSCs) have been studied vigorously \cite{kamihara2008iron,hanaguri2010unconventional,paglione2010high,hirschfeld2011gap,stewart2011superconductivity,wang2011electron,chubukov2012pairing,chubukov2014fe}. A substantial diversity of materials has been realized in this class, all sharing characteristically high critical temperatures and rich phase diagrams with nearby magnetic phases, as a function of chemical substitution. More recently it has been appreciated that several members of this family also have topologically nontrivial normal-state band structures \cite{wang2015topological,xu2016topological,hao2014topological,wu2015cafeas,wu2016topological,hao2018topological,konig2019helical,qin2019quasi}, and therefore may be a natural platform for the realization of Majorana zero modes (MZM) \cite{read2000paired,kitaev2001unpaired,sau2010nonabelian,alicea2012new} and effective topological superconductivity (TSC) \cite{qi2009time,fu2010odd,qi2011topological,zhang2013time}. As in the original Fu-Kane proposal \cite{fu2008superconducting}, the pairing symmetry is taken to be conventional; then, through a ``self-proximity effect'' induced by the bulk superconductor, the topological surface states or vortex lines are imbued with pair correlations from the bulk, leading to lower-dimensional surface or vortex TSC.

This indeed appears to be the case with the vortex core MZM recently observed in the unconventional iron-based superconductor FeTe$_{1-x}$Se$_{x}$ ($x=0.45$) (FTS) \cite{wang2018evidence,zhang2018observation,zhang2018multiple,machida2018zero}. Above the superconducting transition at $T_c= 14.5$K, this system exhibits non-trivial band topology with a band inversion along the $\Gamma-Z$ line in the Brillouin zone. On the (001) surface, angle-resolved photoemission spectroscopy (ARPES) measurements clearly see a helical Dirac surface state \cite{zhang2018observation}, the hallmark of three dimensional (3D) time-reversal invariant (TRI) topological insulators (TI) \cite{fu2007topological,fu2007inversion}. Below $T_c$, the surface state develops a gap, as expected. STM measurements of the same system in a weak magnetic field reveal robust zero-bias peaks inside vortex cores, a strong indication of zero energy subgap states and associated MZM physics \cite{wang2018evidence}.

In this letter, we present a completely different mechanism for Majorana states in FeSCs in the absence of vortices. Specifically, we demonstrate the emergence of higher-order topology \cite{benalcazar2017quantized,zhang2013surface,benalcazar2017electric,schindler2018higher,langbehn2017reflection,khalaf2018higher,liu2018majorana,peng2018proximity,shapourian2018topological,volpez2018second,wang2018high,yan2018majorana,wang2018weak,wu2018higher,zhu2018tunable,bultinck2018three,pan2018lattice,ghorashi2019second} and corresponding helical Majorana hinge states in an explicit minimal lattice model based on two key ingredients \cite{highertopo}: (i) the band inversion along $\Gamma-Z$ \cite{xu2016topological,zhang2018observation} and (ii) extended $s$-wave ($s_{\pm}$) pairing \cite{hanaguri2010unconventional,hirschfeld2011gap,chubukov2012pairing}, with a sign change of the pairing potential between the $\Gamma$ and $M$ points. We expect this to be a reasonable effective model that captures the topological properties of FTS \cite{zhang2018observation,wang2018evidence}, Li$_{0.84}$Fe$_{0.16}$OHFeSe \cite{liu2018robust}, LiFeAs \cite{zhang2018multiple}, and other FeSCs. We explain the origin of these Majorana hinge states with an analytic theory for the projection of the bulk extended s-wave pairing onto the Dirac cone of an arbitrary surface termination, and numerically demonstrate their stability at nonzero chemical potential and in the presence of chemical potential disorder.

\begin{figure}[t]
	\centering
	\includegraphics[width=0.48\textwidth]{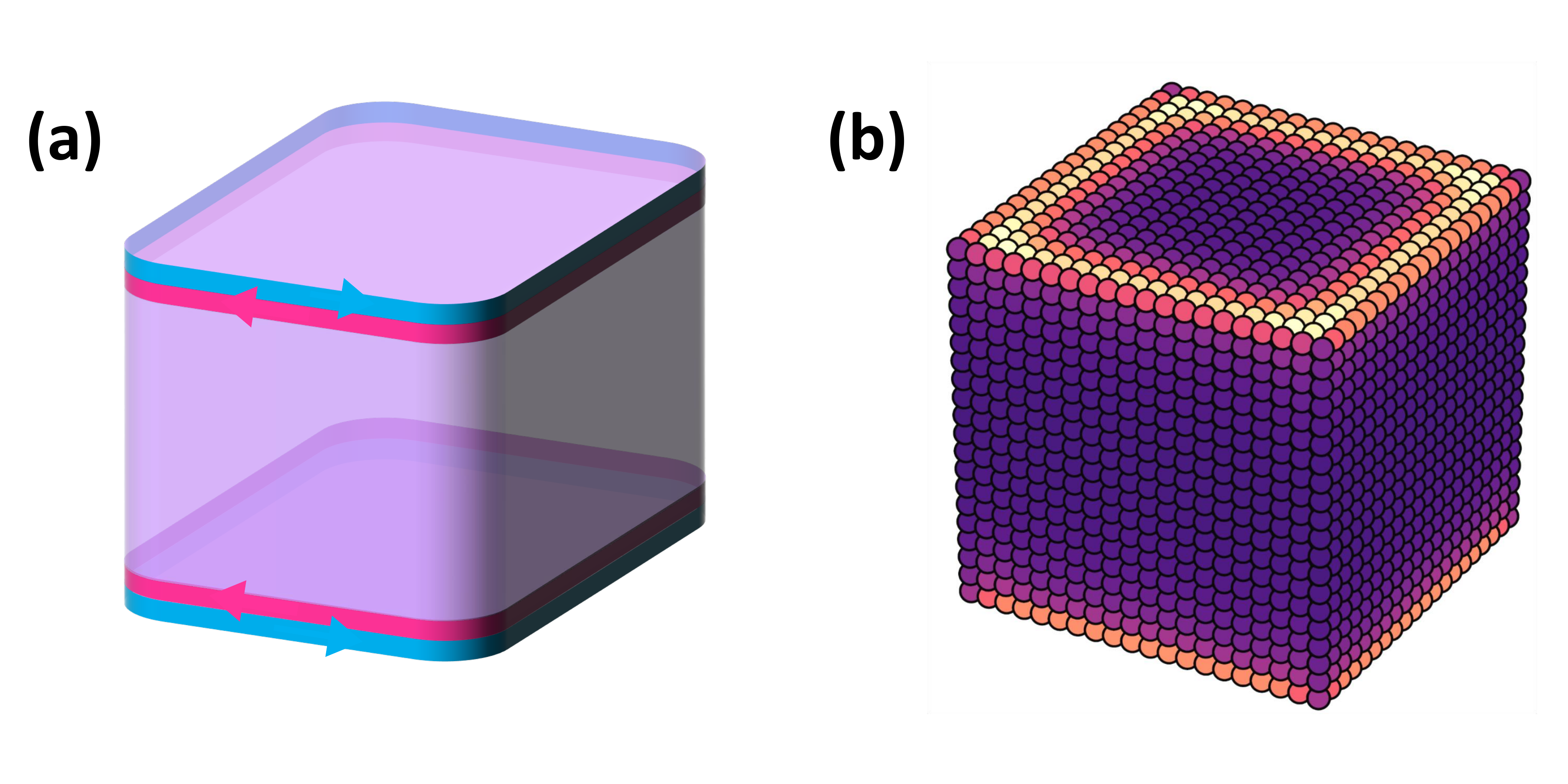}
	\caption{(a) Schematic plot of the helical hinge Majoranas on the edges between top/bottom surfaces and side surfaces. Red and blue contours indicate the left- and right-moving Majorana modes, respectively. (b) $|\psi(r)|^2$ for the lowest-energy eigenstate of the lattice model on a cube geometry with open boundary conditions in all directions has support exponentially localized to the hinges.}
	\label{Fig: schematic}
\end{figure}

{\it Model Hamiltonian} - Our minimal BdG model describes a 3D TI with $s_{\pm}$ pairing on a cubic lattice:
\bea
H(k) = \begin{pmatrix}
	H_0(k)-\mu & -iD(k) \\
	iD(k) & \mu - H_0^*(-k)
\end{pmatrix}.
\eea
The normal state Hamiltonian is $H_0(k)=v(\sin k_x\Gamma_1 + \sin k_y\Gamma_2+\sin k_z \Gamma_3)+ m(k)\Gamma_5$ with $m(k)=m_0-m_1(\cos k_x + \cos k_y)-m_2\cos k_z$ \cite{liu2010model}. The $4 \times 4$ matrices $\Gamma_i$ are chosen to be
\bea
\Gamma_1 &=& \sigma_x \otimes s_x,\ \Gamma_2 = \sigma_x \otimes s_y,\ \Gamma_3 = \sigma_x \otimes s_z, \nonumber \\
\Gamma_4 &=& \sigma_y \otimes s_0,\ \Gamma_5 = \sigma_z \otimes s_0,
\eea
where $\sigma$ and $s$ are Pauli matrices for the orbital and spin degrees of freedom, respectively. The time-reversal ($\Theta$) and the parity ($P$) symmetry operations are given by
\bea
\Theta = i\sigma_0\otimes s_y K = -i\Gamma_{13} K,\ P=\sigma_z\otimes s_0 = \Gamma_5
\eea 
where $\Gamma_{ij}=[\Gamma_i,\Gamma_j]/2i$. The band topology of $H_0$ is easily read off using the Fu-Kane criterion \cite{fu2007inversion}. 
For our purposes, we choose $v=1,m_0=-4,m_1=-2,m_2=1$ which locates the band inversion at $Z$ \footnote{The band inversion at $Z$ happens when $m_0<0, m_1<0, m_2>0, 2m_1-m_2<m_0<\text{min}\{-m_2,-|2m_1+m_2|\}$.}. $H_0$ then describes a strong TI phase with helical Dirac surface states.

The $s_{\pm}$ pairing function is $D(k) = \Delta(k) \Gamma_{13}$, with
\bea
\Delta(k) = \Delta_0 + \Delta_1 (\cos k_x + \cos k_y)
\eea
We first restrict to $\mu=0$ for simplicity, which is also justified by the small chemical potential observed in the ARPES experiments. As we adiabatically turn on the pairing, the bulk remains a topologically trivial superconductor. However, unlike the case of uniform $s$-wave pairing, the $s_{\pm}$ pairing will gap out the TI surface states in an anisotropic way. This is the basic principle that enables the realization of new surface topological phenomena beyond the conventional bulk topology.

\begin{figure}[t]
	\centering
	\includegraphics[width=0.49\textwidth]{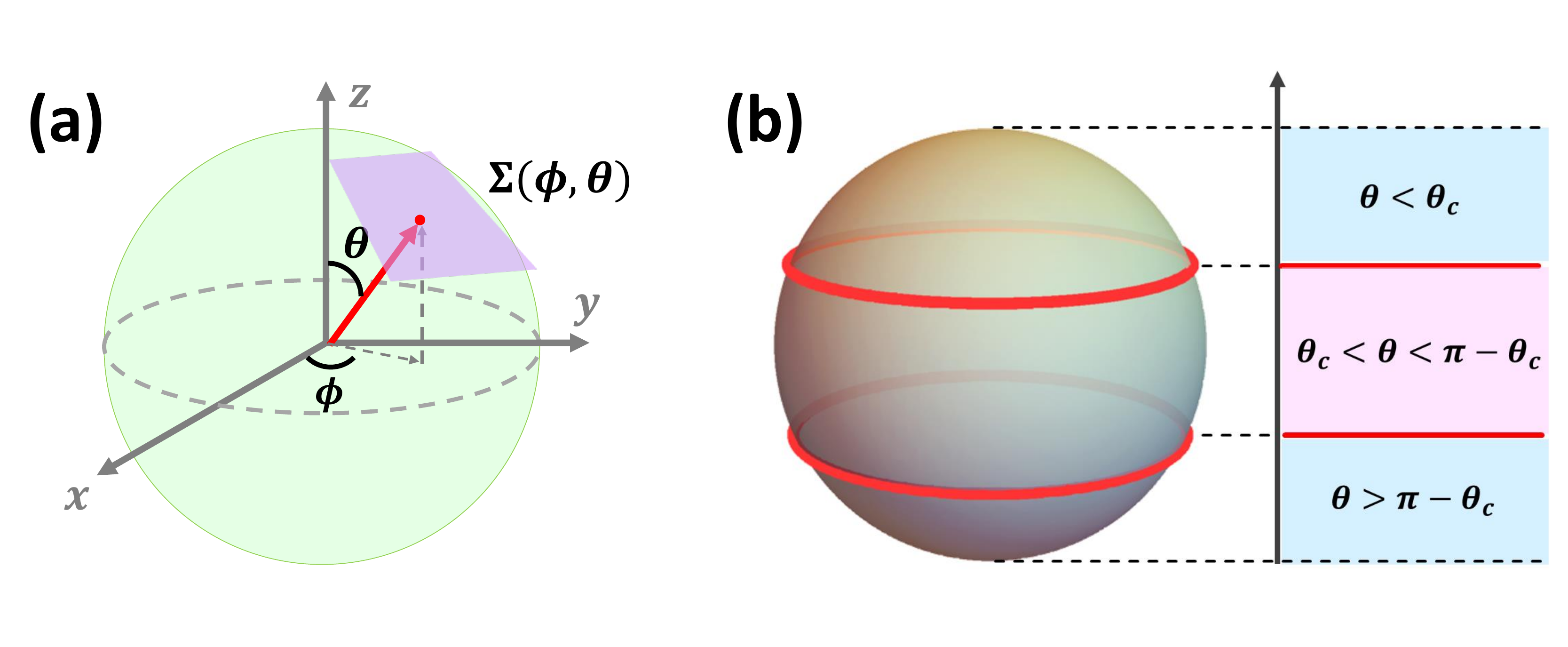}
	\caption{(a) To describe an arbitrary facet $\Sigma(\phi,\theta)$ (purple surface), we consider a spherical geometry and use the Euler angles $\phi$ and $\theta$ to define a normal vector ${\bf n}_{\Sigma}(\phi,\theta)$ (red arrow) that uniquely labels $\Sigma(\phi,\theta)$. (b) A schematic plot of the surface gap evolution as a function of $\theta$ when $\theta_c$ exists. The blue and red color denote surface pairings with different signs. }
	\label{Fig: coordinate}
\end{figure}

{\it Theory of Surface States} - To visualize the anisotropy of surface state pairing, we first notice that any surface termination of a crystal can be described as a tangent plane of the unit sphere \cite{zhang2012surface,zhang2013surface}, as shown in Fig.~\ref{Fig: coordinate}(a). In particular, an arbitrary surface $\Sigma(\phi,\theta)$ can be uniquely labeled by its unit normal vector defined as ${\bf n}_{\Sigma}=(\sin \theta \cos \phi, \sin\theta \sin \phi, \cos\theta)^T$. In the long-wavelength limit, we expand $H_0(k)$ around $Z$ to ${\cal O}(k^2)$ and obtain $H_0^Z({\bf k})=v(k_x\Gamma_1+k_y\Gamma_2-k_z\Gamma_3) +[\tilde{m}_0+\frac{m_1}{2}(k_x^2+k_y^2)-\frac{m_2}{2}k_z^2]\Gamma_5$, where $\tilde{m}_0=m_0-2m_1+m_2$. To solve for the effective surface theory for $\Sigma(\phi,\theta)$, we apply the coordinate rotation $R(\phi,\theta)=R_Y(-\theta)R_Z(-\phi)$ to ${\bf k}=(k_x,k_y,k_z)^T$ and arrive at
\bea
{\bf k}'&=&(k_1,k_2,k_3)^T = R(\phi,\theta){\bf k},
\eea
where $R_Y(\theta)$ and $R_Z(\phi)$ are Euler rotations around $y$-axis and $z$-axis, respectively. It is easy to check that $k_3 = {\bf n}_{\Sigma}\cdot {\bf k}$ and any surface state on $\Sigma(\theta,\phi)$ can now be obtained by imposing open boundary conditions on $H_0^Z({\bf k}')$ along $k_3$ direction.

Generally, $H_0^Z({\bf k}')$ has a complicated form in ${\bf k}'$. However, we notice that there always exists the unitary transformation $U(\phi,\theta)=e^{i\Gamma_{13}\theta/2}e^{i\Gamma_{12}\phi/2}$ such that $\tilde{H}_0^Z=U(\phi,\theta) H_0^Z U(\phi,\theta)^{\dagger}=\tilde{h}_0 + \tilde{h}_1$ has a simple form:
\bea
\tilde{h}_0 &=& -vk_3\Gamma_3 + (\tilde{m}_0 - \tilde{m}_2k_3^2) \Gamma_5 \nonumber \\
\tilde{h}_1 &=& v(k_1\Gamma_1+k_2\Gamma_2)+ (\tilde{m}_{13}k_1k_3+\tilde{m}_1k_1^2+\frac{m_1}{2}k_2^2)\Gamma_5. \nonumber \\
&&
\eea
We have defined $\tilde{m}_1=(m_1\cos^2\theta-m_2\sin^2\theta)/2$, $\tilde{m}_2=(m_2\cos^2\theta-m_1\sin^2\theta)/2$ and $\tilde{m}_{13}=(m_1+m_2)\sin 2\theta/2$. Now we solve for the zero mode equation of $\tilde{h}_0(k_1=k_2=0)$ and consider $\tilde{h}_1$ as a perturbation to extract the surface state dispersion of $\Sigma(\phi,\theta)$. 

Replacing $k_3 \rightarrow -i \partial_{x_{3}}$ yields the zero mode equation
\bea
[iv\partial_{x_3}\Gamma_3 + (\tilde{m}_0+\tilde{m}_2\partial_{x_3}^2)\Gamma_5]\tilde{\psi}(x_3) = 0
\eea
With the boundary conditions $\tilde{\psi}(x_3=0)=\tilde{\psi}(x_3=-\infty)=0$, we find two solutions $\tilde{\psi}_{1,2}$ that are exponentially localized near $x_3=0$:
\bea
\tilde{\psi}_i = {\cal N}\sin (\beta x_3) e^{\alpha x_3} e^{i(k_1x_1+k_2x_2)} \xi_i,\  i=1,2.
\label{Eq: zero mode of TI}
\eea
where we have defined
\bea
\alpha = \frac{v}{2\tilde{m}_2},\ \beta = \frac{\sqrt{4\tilde{m}_0\tilde{m}_2-v^2}}{2\tilde{m_2}},
\eea
and the normalization factor ${\cal N} =\sqrt{4\alpha(\alpha^2+\beta^2)/\beta^2}$. The spinor part of $\tilde{\psi}_i$ are the eigenstates of $\Gamma_{35}$ with eigenvalue $-1$: 
\bea
\xi_1 = (0,i,0,1)^T,\ \xi_2 = (-i,0,1,0)^T.
\label{Eq: TI zero mode_spinor}
\eea
Treating $\tilde{h}_1$ as a perturbation, the low energy description of the surface state is given by
\bea
h_{ss} = k_1\varsigma_2+k_2\varsigma_1,
\eea
where $\varsigma$ are pseudo-spin Pauli matrices in the space spanned by $\xi_1$ and $\xi_2$. 

{\it Surface State Pairing} - The effective pairing $h_{\Delta}(\theta,\phi)$ on a surface $\Sigma(\phi,\theta)$ is solved by treating the $s_{\pm}$ SC pairing $\Delta(k) \tau_y \otimes \Gamma_{13}$ as a perturbation ($\tau$ are the Pauli matrices for the particle-hole basis). The particle-hole redundancy requires the hole counterparts of Eq.~\ref{Eq: zero mode of TI}, and the spinor part of the zero modes in the BdG basis are
\bea
\chi_1 &=& (1,0)^T\otimes U^{\dagger}\xi_1,\ \chi_3 = (0,1)^T\otimes (U^{\dagger}\xi_1)^*, \nonumber \\
\chi_2 &=& (1,0)^T\otimes U^{\dagger}\xi_2,\ \chi_4 = (0,1)^T\otimes (U^{\dagger}\xi_2)^*,
\eea
where we have undone the unitary transformation $U(\phi,\theta)$ of $\xi_i$. By rotating $\Delta({\bf k})$ to $\Delta({\bf k}')$ and expanding it around $Z$, the effective pairing at the surface Dirac point with $k_1=k_2=0$ becomes
\bea
\Delta(0,0,-i\partial_{x_3}) = \Delta_0 + 2\Delta_1 + \Delta_1 \frac{\sin^2\theta}{2}\partial_{x_3}^2.
\eea
We expect that $h_{\Delta}(\phi,\theta)=\Delta_\text{eff}(\theta)\Lambda(\phi,\theta)$, where $\Delta_\text{eff}$ and $\Lambda$ come from the zero mode projection of the scalar momentum part $\Delta(k)$ and the matrix part $\tau_y\otimes \Gamma_{13}$ respectively.
Interestingly, the matrix part $\Lambda$ is independent of the surface $\Sigma(\phi,\theta)$:
\bea
(\Lambda)_{i,j} =  \chi_i^{\dagger} \tau_y\otimes \Gamma_{13} \chi_j = (\tau_y\otimes \varsigma_y)_{i,j}.
\eea
The spatial anisotropy of $h_{\Delta}(\phi,\theta)$ arises completely from $\Delta_\text{eff}(\theta)$. Our main analytic result is an approximate formula for the effective surface pairing,
\bea
	\Delta_\text{eff}
	&=& 
	{\cal N}^2\int_{-\infty}^0 dx_3 \sin (\beta x_3) e^{\alpha x_3} \Delta(-i\partial_{x_3})\sin (\beta x_3) e^{\alpha x_3} \nonumber \\
	 &=& \Delta_0 + 2\Delta_1 - \Delta_1 \frac{m_0-2m_1+m_2}{m_2\cos^2\theta - m_1\sin^2 \theta}\sin^2 \theta.
	\label{Eq: Effective Pairing}
\eea
The effective surface SC theory for any facet $\Sigma(\phi,\theta)$ is finally given by,
\bea
H_\Sigma(k_1,k_2)= \begin{pmatrix}
	k_1\varsigma_2 + k_2\varsigma_1 & -i\varsigma_y \Delta_\text{eff}(\theta) \\
	i\varsigma_y \Delta_\text{eff}(\theta) & -k_1\varsigma_2 + k_2\varsigma_1
\end{pmatrix}.
\label{Eq: surface Hamiltonian}
\eea

\begin{figure}[t]
	\centering
	\includegraphics[width=0.48\textwidth]{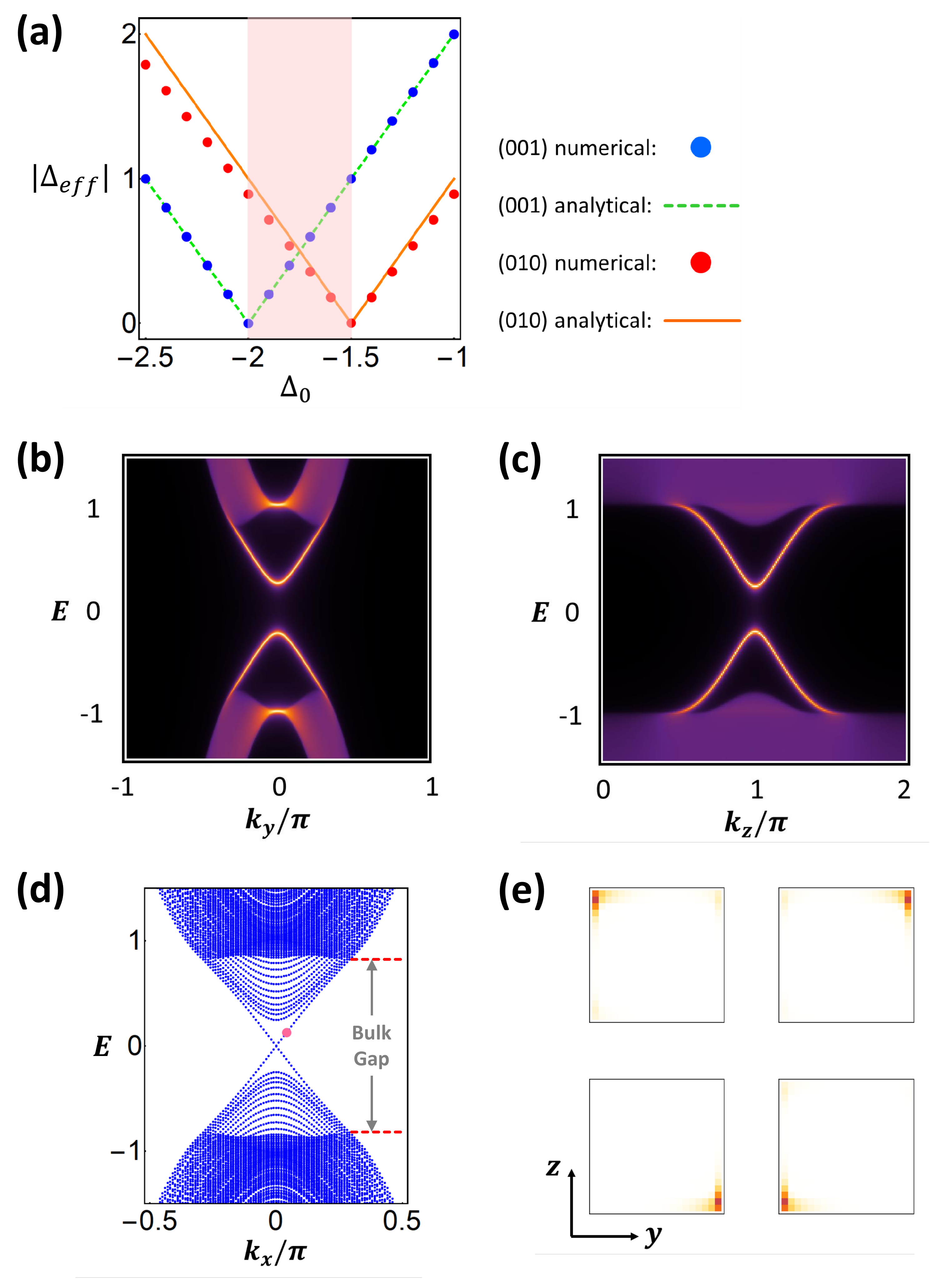}
	\caption{(a) Evolution of the $(001)$ and $(010)$ surface gaps. The green dashed and red solid line use Eq. \ref{Eq: Effective Pairing}. Results for both surfaces agree well with numerical calculations on the full lattice model (dots). The red region denotes the phase with helical hinge Majoranas. In (b) and (c), we plot the surface spectrum of both $(001)$ and $(010)$ surfaces using the iterative Green function method. (d) Energy spectrum in a wire geometry along $x$ with open boundary conditions on both $y$ and $z$ with 20 lattice sites along each direction. The linear modes inside the surface gap are the helical hinge Majoranas. (e) Spatial profile of the eigenstates at the red dot in (d).}
	\label{Fig: sphere}
\end{figure}

{\it Helical hinge Majoranas} - The angular anisotropy of $\Delta_\text{eff}(\theta)$ enables a nontrivial situation where the pairing gap flips its sign while crossing the hinge between two facets $\Sigma(\phi,\theta_1)$ and $\Sigma(\phi,\theta_2)$. Mathematically,
\bea
\Delta_\text{eff}(\theta_1)\Delta_\text{eff}(\theta_2) < 0.
\label{Eq: Domain wall condition}
\eea
With Eq.~\ref{Eq: Domain wall condition}, the hinge physics manifests itself as a TRI topological domain wall problem of 2D BdG Dirac fermions. As a result, the sign reversal of the pairing gaps necessarily binds the hinge with a pair of 1D helical Majoranas. 

Furthermore, the hinge Majorana condition in Eq.~\ref{Eq: Domain wall condition} requires the existence of a topological critical angle $\theta_c$, where $\Delta_\text{eff}(\theta_c)=0$. This critical angle $\theta_c$ is given by
\bea
\cot^2\theta_c = \frac{m_1}{m_2}+\frac{\Delta_1}{\Delta_0+2\Delta_1} \left( \frac{m_0-2m_1}{m_2}+1 \right).
\eea
Helical hinge Majoranas will appear as long as $\theta_c$ exists, i.e., when $\cot^2 \theta_c > 0$. With our choice of model parameters, the condition is
\bea
-2\Delta_1<\Delta_0<-\frac{3}{2}\Delta_1.
\label{Eq: Topological condition}
\eea
In Fig.~\ref{Fig: coordinate} (b), we schematically plot the surface gap distribution for all facets, for parameters with a non-zero $\theta_c$, where the blue and red regions denote positive and negative surface gaps, respectively \footnote{Notice that spatial inversion symmetry will enforce $\Delta_\text{eff}(\theta) = \Delta_\text{eff}(\pi-\theta)$ in our system.}.

We now confirm this continuum theoretical analysis by carrying out a direct numerical solution of the full lattice model, with results shown in Fig. \ref{Fig: sphere}. The gap is obtained in a slab geometry on the $(001)$ surface with $\theta=0$ and the $(010)$ surface with $\theta=\pi/2$. The numerical results shown in Fig.~\ref{Fig: sphere} (a) agree well with the analytic result from Eq. \ref{Eq: Effective Pairing}. As $\Delta_0$ is tuned from zero, a surface topological phase transition first occurs at $\Delta_0= -\frac{3}{2}\Delta_1$ with vanishing surface gap on the $(010)$ surface. The gap then reverses sign between the $(001)$ and $(010)$ surfaces, and the hinge Majorana condition is satisfied. When $\Delta_0=-2\Delta_1$, a second surface topological phase transition occurs, trivializing the surface topology and eliminating the hinge Majoranas. In Fig.~\ref{Fig: sphere} (b) and (c), we show the clear surface gaps at $\Delta_0=-1.75\Delta_1$ for translation invariant $(001)$ and $(010)$ surfaces. In Fig.~\ref{Fig: sphere} (d), we calculate the band dispersion along $k_x$ for a wire geometry with periodic boundary conditions along $\hat{x}$ and open boundary conditions along $\hat{y}$ and $\hat{z}$. Correspondingly, we observe four pairs of helical Majoranas states that appear inside the surface gap. In Fig.~\ref{Fig: sphere} (e), we plot the spatial profile for these states. Each pair of helical Majoranas is exponentially localized around each corner of the $y$-$z$ cross section (i.e., the edge between $(001)$ and $(010)$ surfaces). For a sample with a cubic geometry, and open boundaries in all directions (as shown in Fig.~\ref{Fig: schematic} (a)) the hinge Majoranas circulate the edges between the top/bottom surfaces and the side surfaces. In Fig. \ref{Fig: schematic} (b), we plot the intensity $|\psi(r)|^2$ for the lowest-energy eigenstate (summed over $\sigma, s, \tau$) of the lattice model on this geometry, showing that the support is exponentially localized to the hinges, as expected.

The appearance of helical hinge Majoranas has a topological origin. In the supplementary materials \cite{supplementary}, we study the Wannier bands \cite{king1993theory,yu2011equivalent,benalcazar2017quantized} of $H(k)$ in different slab geometries and find ``helical" surface Wannier bands that characterize the time-reversal-symmetric $\mathbb{Z}_2$ pump of BdG quasi-particles on the surfaces. Such $\mathbb{Z}_2$ pumps also occur in 2D TRI TIs/TSCs and offer a topological picture for the helical edge physics \cite{fu2006time,yu2011equivalent}. Similarly, the helical surface Wannier band spectrum here unambiguously signals the helical hinge Majoranas. 

We emphasize that the helical surface Wannier bands \emph{should not} be interpreted as a surface topological index for any single isolated surface, e.g. the (001) surface. Importantly, any surface Hamiltonian $H_{\Sigma}$ in Eq. \ref{Eq: surface Hamiltonian} is topologically trivial by itself, and the helical hinge Majoranas only arise when a domain wall is formed between neighboring facets. Therefore, the higher order topology here is a combined effect with contributions from all surfaces in a crystal.

\begin{figure}[t]
	\centering
	\includegraphics[width=0.45\textwidth]{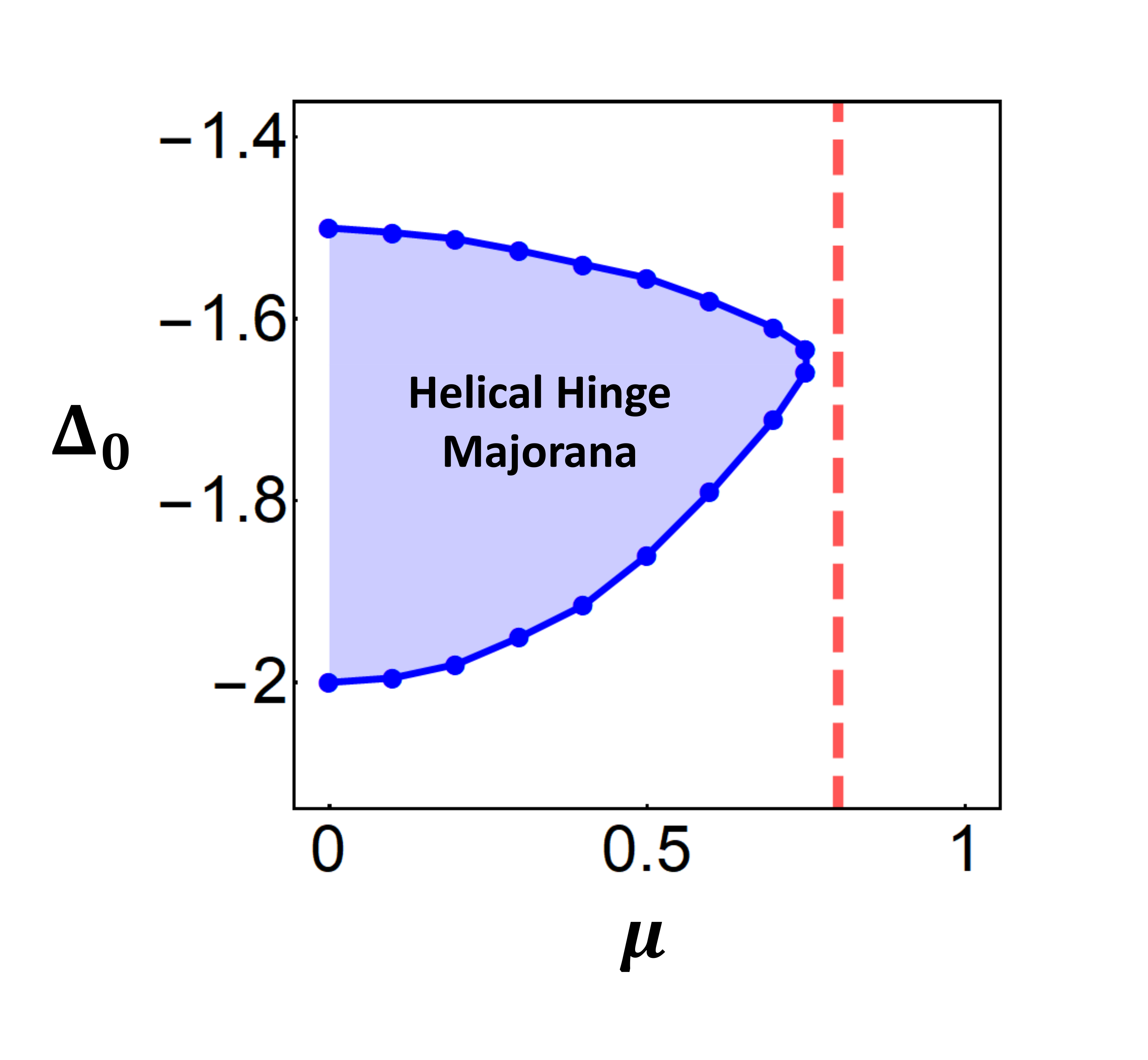}
	\caption{Topological phase diagram in $\Delta_0$ (in unit of $\Delta_1$) and chemical potential $\mu$ (in unit of $m_2$). The bulk conduction band starts at $\mu \simeq 0.8$ (dashed line), which is also where the hinge Majorana phase vanishes.}
	\label{Fig: phase}
\end{figure}

{\it Stability} - Having demonstrated the existence of hinge modes under ideal circumstances, we now show that they persist even when the chemical potential is moved from the Dirac point. The analytic theory can no longer be easily applied for finite chemical potential, but we can characterize the phase hosting hinge modes through either its surface spectrum or its Wannier band spectrum. As the surface gap closes and reopens, the helical (gapped) surface Wannier bands will simultaneously develop a gap (become helical), indicating a surface topological phase transition. By checking both surface energy band spectra and Wannier band spectra, we numerically obtain the phase diagram of our model as a function of the isotropic pairing $\Delta_0$ (in units of $\Delta_1$) and the chemical potential $\mu$, shown in Fig.~\ref{Fig: phase}. We find that the non-trivial phase with helical hinge Majoranas persists, over a decreasing range of $\Delta_0$, until the chemical potential reaches the bulk conduction band. However, this is also the scale at which we expect our effective model to no longer accurately represent the topological bands of realistic FeSCs.

We have also checked the stability against onsite potential variations $V(r) \tau_z \otimes \sigma_0 \otimes s_0$, taking $\left< V(r) \right> = 0$ and $\left< V(r) V(r') \right> = W^2 \delta_{rr'}$. In a real-space calculation using the kernel polynomial method \cite{weisse2006kernel}, we find the hinge states persist for small $W$ (say, half the bulk gap), but defer a full investigation of disorder effects to future work.

{\it Conclusion and Discussion} - Motivated by experimental results on the band topology of several FeSCs, we have constructed an explicit model of higher-order topological superconductivity supporting helical Majorana states at the hinges between different facets, depending on their orientations. The main ingredients are the band inversion at the $Z$ point and $s_{\pm}$ pairing. Unlike some related constructions for 2D higher-order TSC in heterostructures \cite{wang2018high,yan2018majorana,zhu2018tunable,liu2018majorana,zhu2018second,pan2018lattice}, our proposal applies intrinsically for a single FeSC sample, avoiding all the complications of an interface.

Some practical observations are in order. First, the helical hinge Majoranas can be probed by the same STM techniques already used to identify vortex bound states \cite{wang2018evidence}; i.e., the surfaces should be locally gapped everywhere away from the hinges, while the hinge LDOS should be finite and approximately constant at low energies, indicating massless dispersing modes localized to the hinges. In a terraced sample, one expects the two pairs of narrowly separated Majoranas that occur at a terrace edge to hybridize. A terrace edge could still host a massive localized mode below the surface gap, but a more detailed modeling would be required to answer this question fully. Finally, the experimental identification of $s_{\pm}$ pairing remains controversial. While vortex bound MZMs are agnostic to whether the pairing is nodeless or $s_{\pm}$, the appearence of hinge states requires the latter. Therefore, the experimental observation of hinge Majoranas would also be a simultaneous demonstration of $s_{\pm}$ pairing.

In addition to more realistic future numerical modeling of specific FeSCs to search for this physics, we also foresee some interesting generalizations of our theoretical construction. In our model, and in the 2D models of Refs. \cite{wang2018high,yan2018majorana,liu2018majorana}, a first-order TI is proximitized by a nodal pairing function resulting in fully gapped but spatially anisotropic superconductivity; a sign change of the gap between different surfaces creates a domain wall that binds helical Majoranas. We conjecture that a second-order 3D TI with ${s_\pm}$ pairing could realize, under the right conditions, a third-order gapped TSC with corner-localized, 0D Majorana modes. 

{\it Acknowledgement} - R.-X.Z is indebted to Fan Zhang, Fengcheng Wu, Sheng-Jie Huang and Lun-Hui Hu for helpful discussions. R.-X.Z is supported by a JQI Postdoctoral Fellowship. W.S.C. is supported by LPS-MPO-CMTC. This work is also additionally supported by Microsoft Q.

\bibliographystyle{apsrev4-1}
\bibliography{HOTSC}

\onecolumngrid

\subsection{\large Supplemental Material for ``Helical Hinge Majoranas in Iron-Based Superconductors"}

\section{Topological characterization of helical hinge Majoranas}

\begin{figure}[h]
	\centering
	\includegraphics[width=0.9\textwidth]{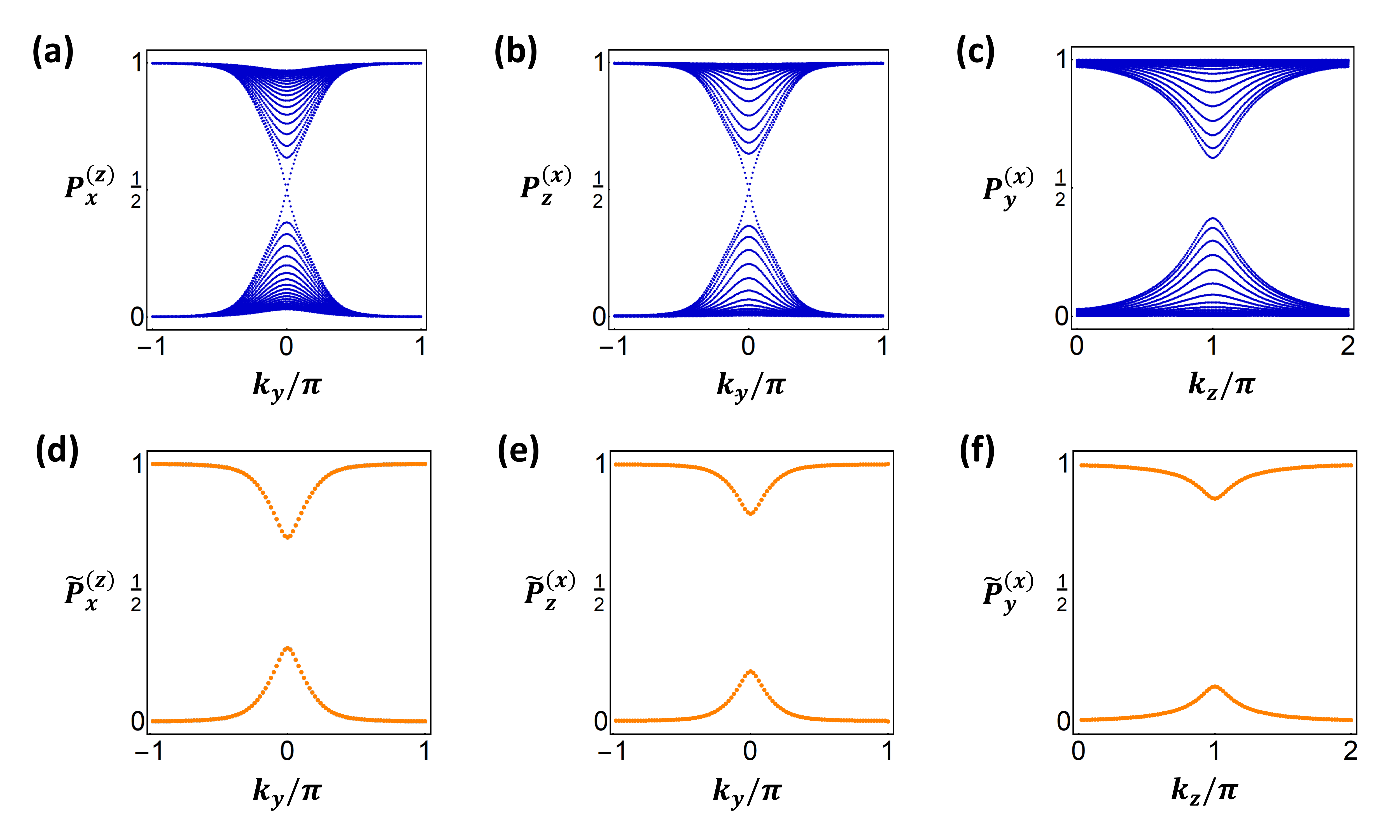}
	\caption{(a) - (c) Wannier bands of the slab Hamiltonian with $20$ slab layers. (d) - (f) Wannier bands of the effective boundary Hamiltonian calculated by the iterative Green function method.}
	\label{Fig: topological}
\end{figure}

In this appendix, we discuss the topological characterization of helical hinge Majoranas in our model using the concept of Wannier bands. Consider a slab geometry where $\hat{x}$ and $\hat{y}$ directions are periodic while $\hat{z}$ direction is open, the corresponding slab Hamiltonian $H^{(z)}(k_x,k_y)$ gives rise to a gapped $(001)$ surface dispersion, as shown in Fig. \ref{Fig: sphere} (b) in the main text. The Wannier center $P^{(z)}_x(k_y)$ at $k_y$ can be calculated by constructing the Wilson loop operator of all occupied BdG bands of $H^{(z)}(k_x,k_y)$ along $k_x$. 

As shown in Fig. \ref{Fig: topological} (a), the evolution of Wannier centers along $k_y$ forms a set of Wannier bands with interesting features: (i) Wannier bands from bulk states are gapped; (ii) Wannier bands from $(001)$ surface states (for both top and bottom surfaces) are gapless. In particular, the ``helical" surface Wannier band of $P^{(z)}_x(k_y)$ characterizes a $\mathbb{Z}_2$ pump of BdG quasi-particles along $y$ direction. Similar $\mathbb{Z}_2$ pump physics has been observed in 2D TRI topological insulators/superconductors and accounts for their heliacl edge physics. Therefore, the helical surface Wannier band is the topological origin of the helical hinge Majorana that circulates the top/bottom surface.

On the other hand, we also plot the Wannier bands of $P^{(x)}_z(k_y)$ and $P^{(x)}_y(k_z)$ in Fig. \ref{Fig: topological} (b) and (c) with open boundary conditions along $\hat{x}$: (i) $P^{(x)}_z(k_y)$ has the same in-gap helical surface Wannier band spectrum as that of $P^{(z)}_x(k_y)$; (ii) $P^{(x)}_y(k_z)$ is gapped. Therefore, there is no non-trivial BdG quasi-particle pumping process along $\hat{z}$ direction, which implies the absence of any in-gap Majorana physics on the edge between $(100)$ and $(010)$ surfaces. This is consistent with our analytical analysis and the numerical results in Fig.~\ref{Fig: schematic} (b). 

However, it should be emphasized that the nontrivial $\mathbb{Z}_2$ pump (helical surface Wannier bands) should not be interpreted as the topological invariant of any single isolated surface system. This is because the effective surface Hamiltonian in Eq. \ref{Eq: surface Hamiltonian} is topologically trivial for any surface $\Sigma$. To prove the above argument numerically, we use iterative Green function method to calculate the boundary Green function $G^{(z)}(\omega, k_x,k_y)$ of the $(001)$ surface with a semi-infinite slab geometry along $\hat{z}$ direction. Unlike the slab Hamiltonian method, the boundary Green function is only concerned with the $(001)$ surface states and does not know anything about the side surfaces. Following Ref. \cite{peng2017boundary}, we define an effective boundary Hamiltonian $H_\text{Green}^{(z)}(k_x,k_y)$ as the inverse of $G^{(z)}(\omega, k_x,k_y)$ at zero frequency $\omega=0$. We further denote $\tilde{P}^{(z)}_x(k_y)$ as the Wannier center of $H_\text{Green}^{(z)}(k_x,k_y)$. As shown in Fig. \ref{Fig: topological} (d), we find the Wannier bands of $\tilde{P}^{(z)}_x(k_y)$ is gapped, which is in contrast to the gapless Wannier bands of $H^{(z)}(k_x,k_y)$ in Fig. \ref{Fig: topological} (a). Similar gapped Wannier bands are also observed for $\tilde{P}^{(x)}_z(k_y)$ and $\tilde{P}^{(x)}_y(k_z)$. This confirms the topological triviality of the effective surface Hamiltonians.

\end{document}